\documentclass[onecolumn,12pt,draftclsnofoot,a4paper,peerreview]{IEEEtran}

\usepackage{amsmath}
\usepackage{amssymb}
\usepackage{bm}
\usepackage{dsfont}
\usepackage{float}
\usepackage{graphicx}
\usepackage{mathrsfs}
\usepackage{psfrag}
\usepackage{subfigure}
\usepackage{times}
\usepackage{xspace}
\usepackage{topcapt}
\usepackage{url,hyperref}

\IEEEoverridecommandlockouts                              
\graphicspath{{figures/}}

\usepackage{color}

\def\baselinestretch{1.5}

\begin{document}

\title{Teaching Wireless Sensor Networks: \\ An Holistic Approach Bridging \\ Theory and Practice at the Master Level}
\author{ \parbox{3 in}{\centering Carlo Fischione\\
         Automatic Control Department\\
         Electrical Engineering and ACCESS\\
         KTH Royal Institute of Technology\\
         10044, Stockholm, Sweden\\
         {\tt\small carlofi@kth.se}}}
\vspace{-0.8cm}
\maketitle 
\vspace{-0.8cm}
\begin{abstract}
Wireless Sensor Networks (WSNs) are a new technology that has received a
substantial attention from several academic research fields in the last years. There are many applications of WSNs, including environmental monitoring, industrial automation, intelligent transportation systems, healthcare and wellbeing, smart energy, to mention a few. Courses have been introduced both at the PhD and at the Master levels. However, these existing courses focus on particular aspects of WSNs (Networking, or Signal Processing, or Embedded Software), whereas WSNs encompass disciplines traditionally separated in Electrical Engineering and Computer Sciences. This paper gives two original contributions: the essential knowledge that should be brought in a WSNs course is characterized, and a course structure with an harmonious holistic approach is proposed. A method based on both theory and experiments is illustrated for the design of this course, whereby the students have hands-on to implement, understand, and develop in practice the implications of theoretical concepts. Theory and applications are thus considered all together. Ultimately, the objective of this paper is to design a new course, to use innovative hands-on experiments to illustrate the theoretical concepts in the course, to show that theoretical aspects are essential for the solution of real-life engineering WSNs problems, and finally to create a fun and interesting teaching and learning environments for WSNs.  
\end{abstract}

\vspace{-0.8cm}
\newpage

\section{Introduction} \label{sec:introduction}

Wireless Sensor Networks (WSNs) are an area of research that has started about fifteen years ago in the academic community of Electrical Engineering and Computer Sciences. A WSN is a network composed of small nodes communicating for actuation, control, and monitoring purposes. WSNs are considered the enabling technology for the internet of things, whereby every physical object equipped by a wireless sensor, for example a human organ, a freezer, a camera, or a jacket, can be connected to the internet~\cite{ZhaoGuibas04}~--\nocite{KarlWillig05,PottieKaiser05,Swami07,Zurawski09,ShelbyBormann09,HykinLiu10,AkyildizVuran10,DargiePoellabauer10,Ferrari10,NikoletseasRolim10}~\cite{Mazumder11}.


Many companies, industrial sites, shops, network operators, need Electrical Engineering and Computer Sciences graduates whom are able and skilled to plan, deploy, and manage WSNs~\cite{ZhaoGuibas04}\nocite{KarlWillig05,PottieKaiser05}\nocite{Swami07,Zurawski09,ShelbyBormann09,HykinLiu10,AkyildizVuran10,DargiePoellabauer10,Ferrari10,NikoletseasRolim10,ChenNixonMok10,WagnerWattenhofer10}~--~\cite{Mazumder11}. Despite a huge research activity on WSNs in the last years, WSNs courses are being introduced in the Master programs only recently and tend to focus on some particular traditional aspects of Electrical Engineering or Computer Sciences, such as Networking, or Signal Processing, or Software Engineering, whereas WSNs is an interdisciplinary area that demands an holistic view of many such topics that are often taught separately. Existing WSNs courses are mostly given at the PhD level, where their structures are hard to digest for Master students. As a result, there is not yet a general approach for the design of WSNs courses at the Master level, the teaching may be difficult, and even the learning can be poor.


The main challenge that we encounter when teaching WSNs is the technical complexity of the subject matter and its interdisciplinary. In WSNs, nodes may have reduced communication, computational, and battery capabilities. They can be mobile device and have reduced coordination. 
When closing the control loops by WSNs, packet dropouts and delays due to retransmissions, channel contentions among transmitters, are typical problems. The software itself, that runs on hardware platforms of small sizes, could introduce unexpected delays or behaviors in the executions that may hinder the estimation, control and actuation operations~\cite{FischionePark+11}. However, in existing Master courses, topics such as Networking, Signal Processing, Controls and Embedded software are taught separately. For example, students in Controls are often thought to design controllers that do not consider the impact of WSNs networking protocols. On the other side, students in Networking are often taught to design WSNs protocols to maximize the successful probability of packet delivery and minimize the delays. This is inefficient because estimators and controllers can tolerate some losses and delays that are different from what is demanded to traditional networks. The WSN networking protocols do not allow to send information to the estimators or controllers at desired times. A major drawback of existing teaching approaches is that the interplay among the typical dynamics introduced by network protocols and estimation and control applications have not been considered. In addition to this, in a WSNs course, students can actually implement code on real devices, design the networking protocol stacks, estimation, control and action functionalities all together, which is not possible in other Controls or Networking or Signal Processing courses. 

The purpose of this paper is to establish a pedagogical methodology to tech WSNs at the Master level for both Electrical Engineering and Computer Sciences students.  
In particular, we aim 1) to provide a systematic approach on how to design a WSNs course, and 2) based on such an approach, to build a course including, a) lectures, b) exercises, c) labs, d) homework, e) exam. The study aims at defining the essential knowledge for WSNs, by reviewing existing papers on how to teach WSNs, and WSNs books. Then, based on these, a WSNs course suitable for a curriculum in Electrical Engineering and Computer Sciences is proposed. What is the relevant knowledge that the students of a WSN course will have to acquire, and how to structure the course so to maximize the learning, is proposed. To achieve the purpose of this paper, the methodology includes data collected from existing course, books, interviews with university teachers, relevant WSNs industrial representatives, and my own experience as a researcher in WSNs. Achieving this purpose appears challenging because there is no adequate research paper, to the best of our knowledge, on how to teach WSNs on both Electrical Engineering and Computer Sciences, as it will be shown by literature review in the related section below.

The rest of the paper is organized as follows. In Section~\ref{literature}, a survey of the existing literature is given. In Section~\ref{relevant knowledge}, the relevance knowledge for a WSNs course is investigated. Based on the results of these two sections, a proposal for a course design and its evaluation is given in Section~\ref{course design}. Finally, the paper is concluded in Section~\ref{conclusions}.

\section{Methodological literature survey} \label{literature}

A first step to develop successfully a course consists in making a literature survey of existing approaches for WSNs (``how to teach'' WSNs), and a survey of the WSNs courses and books (``what to teach'' in the detail). In this section, we highlight potentials and drawback of the existing teaching approaches from papers that discuss how to teach WSNs. The survey of courses and books will be carried out later in the next section. 

The scientific and technical research on WSNs is vast, if not the largest in Electrical Engineering and Computer Sciences in recent years. There are not many works  addressing the need of teaching WSNs, as pointed out in~\cite{Rollins11}. One of the earliest papers on sensor network teaching is~\cite{1013877}, where the authors present a methodology that uses WSNs to introduce students to the concepts of remote team design and systems engineering. The paper advocates the importance of an hands-on approach on WSNs, but since then, many new topics have been developed and proposed for WSNs. Most of the contributions on how to teach WSNs can be found in system-level Computer Sciences~\cite{Rollins11}~--\nocite{Feonster+09,Tyman+09,Feonster+10,Zennaro+12,6375774}~\cite{5598548}, whereas a minor number of works can be found in Electrical Engineering~\cite{Taslidere+11,4909475}. A web group also created an emailing list on WSNs teaching~\cite{teaching-sensornet-website}. Most of these works highlight the importance of hands-on approach for WSNs, which has been first recognized in~\cite{1013877} and later convincingly shown in \cite{Feonster+09} \nocite{Tyman+09}~\cite{Feonster+10}.

The teaching approach suggested in~\cite{Rollins11} proposes that the students are asked to read research articles beforehand and then these articles be discussed in the class room. In~\cite{Rollins11}, a networking perspective for Computer Sciences students is adopted and the authors pay more attention on definition of software primitives for networking, rather than theoretical tools to design the networking aspects of WSNs. The students of the course analyzed in~\cite{Rollins11} have spontaneously proposed  control and robotic applications as an interesting assignment for a future edition of the course, but as the author writes ``this assignment [control and robotic] does not integrate many of the distributed and network concepts that students learn in the [present] course''. Robotic and WSNs teaching is considered in~\cite{6375774,4914744}, where however the focus is on some interesting experiment design within Embedded Systems, whereas Networking and Signal Processing topics are not covered. The teaching requirements for a virtual wireless sensor network laboratory between Greece and USA are described in~\cite{4084615}, where the teaching content is taken from~\cite{ZhaoGuibas04}, and where the authors describe the software interfaces that allow the interaction between Carnegie Mellon University, USA and Athens Information Technology, Greece. 

The articles~\cite{Taslidere+11,4909475} are two of the few ones so far existing on WSNs teaching in an Electrical Engineering context.  However, they focus directly on the design of experimental set-ups to link some of the WSNs theory to real-world engineering applications, without defining first the essential knowledge and learning outcome for WSNs in general. Specifically, in~\cite{Taslidere+11} three experiments are designed for students so as to learn in the areas of wireless embedded networks,
detection and estimation theory, stochastic processes, probability theory, statistical pattern recognition, and digital signal processing. Statistical evidence of the effectiveness of the experiments is therein given, showing that students knowledge after the theoretical lecture and before the hands-on experiments is greatly enhanced  after the experiments. In~\cite{4909475}, a software interface that enables signal processing applications for WSNs is presented. The paper supports the importance of signal processing for WSNs, and then focuses on a proposal for a software interface that can be effective
in explaining both theory and applications of signal processing in WSNs. 

The need of interdisciplinary teaching in WSNs was emphasized in~\cite{5598548}, where a detailed description of course content and tests is provided. However, the course is intended for medical and software/computer engineering and therefore the teaching needs such as Networking, Signal Processing, and Controls are not addressed. In~\cite{Rollins11}, the learning needs for Electrical Engineering, such as how to choose a networking protocols including modulation, coding, and access formats, and how to monitor or control the physical phenomena being sensed, are not considered. The interesting approach proposed in~\cite{Feonster+10} has the drawback of being focused on WSNs from a software engineering point of view. The course therein proposed lacks essential aspects such as important parts of Networking, pertaining to the physical layer of the wireless communication, most of Signal Processing concepts including data analysis, and Controls.

The proposed approaches to ``how to teach" in the papers above are focused on some specific aspect of WSNs and lack the holistic view that should be used in WSNs interdisciplinary area. They are intended for some specific disciplines, such as for example on Networking and Distributed Systems from the point of view of software engineering in Computer Sciences. ``What to teach'' in WSNs for both Electrical Engineering and Computer Sciences does not seem to be addressed. What is missing is a coherent and the interconnected material selection for WSNs. As a result, what students may get from existing courses is something taken here and there, which is difficult to illustrate in a cohesive manner. The students may have difficulty at realizing the interconnection between the important different topics that are present in WSNs, which can in turn minimize dramatically the learning potentials. Based on this literature survey, we conclude that 1) there is no adaptable approach on how to teach a WSN course for both Electrical Engineering and Computer Sciences, and 2) the hands-on approach seems to be the best to understand the complex nature of WSNs and its interaction with the cyberphysical world. 

In the sequel, by filling the gaps in existing teaching approaches, we survey courses and books to investigate ``what to teach" in a WSN course.

\section{Essential knowledge analysis} \label{relevant knowledge}

In this section we have conducted an analysis to pinpoint the essential knowledge that concurs to the definition of a WSN course. The analysis consists in a taxonomy of the essential topics based on relevant existing courses, important existing books, faculty members, students, industrial representatives interviews, and personal experience. 

The courses under analysis are reported in Table \ref{tab:existing-courses}. The choices of the courses is based on the following criteria: the instructors are leading scientists in the area, or the university are leading teaching centers, or the instructors have written pedagogical research papers on WSNs mentioned in Section~\ref{literature}. The books we have considered are reported in the citation list below as~\cite{ZhaoGuibas04}--\cite{Mazumder11}. The selection of these books is based on the scientific reputation of the authors. 

\begin{table}[t]
\tiny
\begin{center}
\caption{Some relevant existing courses. ``'M'' stands for Master course, ``'PhD'' stands for PhD course. The weblinks were checked on August 2013}
  \begin{tabular}{|l ||  l || l || l  ||}
     \hline Label & Institution & Title and web address& Course instructor\\
     \hline 
     \hline 1 & Harvard, USA & Wireless Sensor Network, PhD & Matt Welsh \\
                &                       & \url{http://www.eecs.harvard.edu/~mdw/course/cs263/} & \\
     \hline 2 & Stanford University, USA & Sensor Network Systems, PhD &Phil Levis \\
                 &                                      & \url{http://www.stanford.edu/class/cs344e/} & \\
     \hline 3 & UCLA, USA & Undergraduate Sensing Courses, M & -- \\
       		 &                    & \url{http://research.cens.ucla.edu/education/undergraduate/courses.htm} & \\
     \hline 4 & UCLA, USA & Graduate Sensing Courses, PhD & --\\
                 &                   & \url{http://research.cens.ucla.edu/education/graduate/courses.htm} & \\
     \hline 5 & KTH, Sweden & Programming WSNs: A system perspective, PhD & Adam Dunkels, Olaf Landsiedel \\
                                                  &  & \url{http://www.kth.se/student/kurser/kurs/EL2747?l=en} & Luca Mottola \\
     \hline 6 & University of San Francisco, USA &  Wireless Sensor Networks, PhD & Sami Rollins \\
                & & \url{https://sites.google.com/site/usfcs685/} & \\
     \hline 7 & University of Nebraska, USA &  Sensor Networks, PhD & M. Can Vuran \\
                 & & \url{http://cpn.unl.edu/?q=wsn} & \\
     \hline 8 & ETHZ, Switzerland & Ad Hoc and Sensor Networks, PhD & Roger Wattenhofer \\ 
                 & & \url{http://www.dcg.ethz.ch/lectures/asn/} & \\
     \hline 9 & Ege University, Turkey & Wireless Sensor Networks, PhD & Kayhan Erciyes\\
                & & \url{http://ube.ege.edu.tr/~erciyes/UBI532/} & \\
     \hline 10 & Stonybrook University, USA & Algorithms for Wireless Sensor Networks, PhD & Jie Gao\\
                  &                                            & \url{http://www.cs.sunysb.edu/~jgao/CSE590-spring11/} & \\
     \hline 11 & National Taiwan University, Taiwan & Wireless Sensor Network And Laboratories, M & Polly Huang\\
                  &                                            & \url{http://nslab.ee.ntu.edu.tw/courses/wsn-labs-fall-10/} & \\
     \hline 12 & University of Lugano, Switzerland & Introduction to Wireless Sensor Networks, M &  Anna F\"orster \\
                  &                                            & \url{http://www.dti.supsi.ch/~afoerste/downloads/WSNCourse2008.zip} & \\
     \hline 13 & University California, Berkeley & Introduction to Wireless Sensor Networks, PhD &  Kristopher Pister, Thomas Watteyne  \\
                  &                                            & \url{http://www.eecs.berkeley.edu/~watteyne/290Q/index.html} & \\
                  
       \hline
  \end{tabular}
  \label{tab:existing-courses}
\end{center}
\end{table}

\subsection{Relevant topics classification} \label{subsec:topic-classific}

Based on the course and book lists, we propose an original classification as reported in Table~\ref{tab:topics}. Each of the entry of the table should be considered a basic bit of the essential knowledge. Given the different nomenclature or meaning intended for the topics, we describe them in the following. Such a description is instrumental  to structure a course, as we will propose later, and because terms may have not a common definition
among disciplines. For example, in the Computer Sciences paper~\cite{Rollins11}, ``Networking''
is often referred to the software primitives that allow the implementation of networking communication, whereas in Electrical
Engineering the term is usually referred to the theory of networks, which is much different from software components. 

\begin{table}[t]
\tiny
\begin{center}
\caption{WSNs relevant topics: name and brief description, as from the courses in Table~\ref{tab:existing-courses} and books~\cite{ZhaoGuibas04}--\cite{WagnerWattenhofer10}.}
  \begin{tabular}{||  l || l  || l  ||  }
   \hline Number & Name & Topics \\
   \hline
    \hline 1 & Antennas & Study of the antennas suitable for WSNs \\
    \hline 2 & Applications  &  Building Automation, Monitoring, Healthcare, Intelligent Transportation Systems, etc. \\
    \hline 3 & Classification & Essentials of machine learning for classification of WSNs information \\
    \hline 4 & Controls  & How to control the protocols and design control applications over WSNs\\
    \hline 5 & Cross layer optimization & Optimization methods in cross layer interaction \\
    \hline 6 & Detection  & Essentials of detection theory for typical WSNs applications \\
    \hline 7 & Estimation  & Essentials of estimation theory for typical WSNs applications \\
    \hline 8 & Hardware platform  & The main components of a hardware platform \\
    \hline 9 & Information processing  & Aggregation and compression of information \\
    \hline 10 & Information theory  & Theoretical tools for detection, source and channel coding \\
    \hline 11 & Localization and positioning & How to locate nodes and objects by a WSN \\
    \hline 12 & Medium access control protocols  & Study of low data-rate and low power MAC \\
    \hline 13 & Modulations  & Study of essential aspects of modulation theory with focus on energy efficiency \\
    \hline 14 & Network management  & Management of the network topology and operation \\
    \hline 15 & Operating systems  & Study of popular operating systems \\
    \hline 16 & Programming  & How to program resource constrained sensors \\
    \hline 17 & Radio propagation  & Study of the wireless channel \\
    \hline 18 & Routing protocols  & Study of low data-rate and low power routing protocols \\
    \hline 19 & Sensor principles  & Physical principles that translates phenomena into electrical signals \\
    \hline 20 & Synchronization  & How to synchronize nodes of a WSN \\
    \hline 21 & Security and Privacy & Elements of secure MAC, routing, modulation and applications \\
    \hline 22 & Standard protocols  & Study of Standards such as IEEE 802.15.4, RPL, Zigbee\\
     \hline 23 & Transport and Appl. Layers, Internetworking  & Study of the Transport and Application OSI layers\\
    \hline
  \end{tabular}
  \label{tab:topics}
\end{center}
\end{table}

The entries of Table~\ref{tab:topics} are specified as follows in alphabetical order: 

{\em 1) Antennas}: pertains to the study of the properties of the antennas used to transmit and receive signals, with particular focus on low power and low data rate devices, because they are the typical ones used for WSNs. The radiation diagram of the antenna should be characterized so to understand how signals are transmitted and received over the spatial directions. 

{\em 2) Applications}: is about the study of the typical applications of WSNs. For example, the application in smart buildings, networks of sensors are used to track people or to help with the activation of the air ventilation and cooling. Another example is the use of sensors for micro smart grids, where the sensors can be connected with the price fluctuations of the electricity and activate the electronic appliances when such prices are low. The list of application of sensor networks is huge, and it is therefore hard to give a comprehensive overview. However, the main categories should be identified and examples for each category should be given in a course. We believe that these categories are the following:
\begin{itemize}
\item Environmental monitoring, where WSNs are used for monitoring of earthquakes, volcanos, fields, seas, lakes and rivers;
\item Industrial automation, where WSNs are deployed to remove communication cables and make more flexible the automation and control of processes;
\item Information and communication technologies, where WSNs are used for telecommunication services;
\item Intelligent transportation systems, where WSNs are used to assist the driving, the drivers, and the traffic;
\item Healthcare and wellbeing, where WSNs are used for monitoring and controlling the rehabilitation and patients, and for training or sport applications;
\item Smart energy, where WSNs are deployed to help with the reduction of the energy consumption in smart cities and smart buildings.
\end{itemize}

{\em 3) Classification and learning}: consists in the study of the basic mechanisms to classify, making inference and prediction from data that have been collected by sensor nodes, including learning theory methods. Classification basically consists in finding the boundaries among classes of data collected by the sensors. 

{\em 4) Controls}: is about basic control theory tools that are necessary to design automatic control applications over WSNs. In particular, the proportional, integrator and derivative controllers are included together with their performance with respect to the networking aspects, such as message losses and message delays introduced by the network protocol.The control of the networking protocols is a part of this topic. 

{\em 5) Cross layer optimization}: deals with the optimization of interaction mechanism among the layers of the protocols. The basics of convex optimization are included so to model mathematically the interactions and to solve typical optimization problems where the goal is to design efficient WSNs respecting the physical constraints to model the delay and packet losses from the communication.

{\em 6) Detection}: deals with the essential theoretical design tools to perform detection of signals by WSNs. The focus is on distributed detection schemes, where the networking aspects play an important role.  

{\em 7) Estimation}: is about the essential theoretical design tools to perform estimation of parameters associated to the signals detected by WSNs. The focus is on distributed estimation schemes, where the networking aspects play an important role.  

{\em 8) Hardware platform}: the main components of a hardware platform that makes up a sensor node are included, considering the node architecture with the memory, sensing, processing, and transmitting components. 

{\em 9) Information processing}: deals with the techniques to process information collected by the nodes such as quantization, compression, and aggregation so to reduce the information to transmit. 

{\em 10) Information theory}: includes a selection of the essential
theoretical tools from general information theory, which are necessary to make efficient source and channel coding with particular reference to constrained sensor nodes.

{\em 11) Localization and positioning}: pertains to signal processing techniques to locate nodes and objects by a WSN and modeling of the physical sources to perform localization such as gyroscopes and accelerometers. Classic techniques such as triangulation and least squares method are part of this topic. The effect of highly noisy measurements provided by the sensors and the short range communication are  included. 

{\em 12) Medium access control protocol}: considers the study and classification of protocols at the medium access control (MAC) level, namely how to make the transmission of messages when multiple nodes try to make transmissions. 

{\em 13) Modulations}: is about the digital modulation formats that are available to shape the transmission of information over the communication channel. The focus is on energy efficiency and the possibility to select the modulation formats in agreement with the requests of the other communication layers. 

{\em 14) Network management}: deals with the problems of deciding the location of nodes and managing their topology so that quality of services of the communication are satisfied. For example, how to place the node so to ensure a desired monitoring of an area, and how to design duty-cycling techniques. 

{\em 15) Operating systems}: includes the most popular operating systems used by WSNs nodes. For example, TinyOS and Contiki. One of these operating systems should be selected to illustrate the application of the theoretical topics in the WSN course. 

{\em 16) Programming}: is about how to program the functionalities of resource constrained sensor nodes by an operating system. Usually, it requires the basic of C programming language. For example, how to encode the automatic selection of modulation formats, medium access control parameters, and routing. How to encode an estimator, or how to connect a plant or process being controlled to a controller running on top of the application layer of the operating system are additional examples.

{\em 17) Radio propagation}: covers the characteristic of the wireless channel, namely how the transmitted power is attenuated. Popular channel fading models such as Rayleigh fading, Ricean fading, Nakagami fading, Log-Normal fading, as well as AWGN channels, are part of this topic. 

{\em 18) Routing protocols}: it studies the low data-rate and low power routing protocols for WSNs. Classic fundamental algorithms such as Ford have to be covered. The topic includes solutions proposed in Standards, such as the Internet Engineering Task Force Routing over Low Power and Lossy Networks. In addition, optimal distributed mechanisms to select the route are included. 

{\em 19) Sensor principles}: is a topic concerning the physical principles that translates sensed phenomena into electrical signals. The mathematical modeling of the signals, along with the characterization of the noises is included for popular sources of signals, such as humidity, pressure, temperature and sounds, and the statistical modeling of the measurement noises. 

{\em 20) Synchronization}: is referred to the techniques to synchronize the clocks of the nodes of a WSN. The basic techniques, along with the performance characterization, are considered. 

{\em 21) Security and privacy}: refers to the techniques that make the MAC, routing, modulation and applications secure with respect to attacks from malicious nodes in a WSN. 

{\em 22) Standard protocols}: includes the study of the most popular standards for WSN networking, such as IEEE 802.15.4 for the physical and MAC layer, and Internet Engineering Task Force Routing over Low Power and Lossy Networks for the routing. The focus is on the mechanisms of the Standard that allow to implement MAC and routing protocols.  

{\em 23) Transport and application layer and internetworking}: includes the OSI protocol levels of transport and application, and how to connect a WSN and its nodes to internet. 

Given the previous classification of the essential WSN topics, it is useful to examine the relative emphasis placed on these topics by the considered courses tabulated in Table~\ref{tab:existing-courses} and books~\cite{ZhaoGuibas04}~--\nocite{KarlWillig05,PottieKaiser05,Swami07,Zurawski09,ShelbyBormann09,HykinLiu10,AkyildizVuran10,DargiePoellabauer10,Ferrari10,NikoletseasRolim10}~\cite{Mazumder11}, in order to get insights into the existing resource structures and to identify their teaching gaps.This is the focus of next subsection.

\subsection{Topic frequency of occurrence}

\begin{figure}[t]
\centering
\includegraphics[width=9cm]{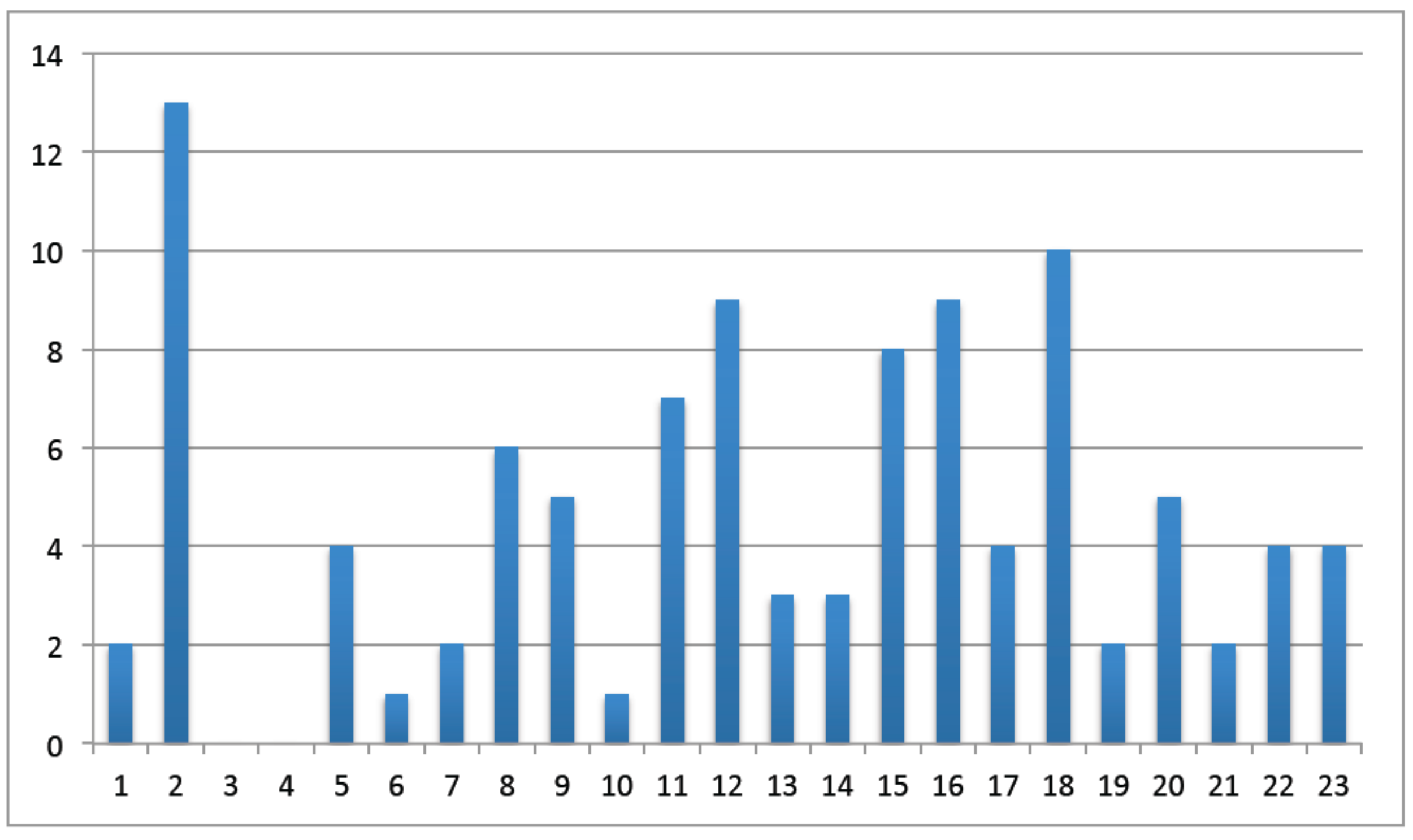}
\caption{Number of occurrences of a topic of Table~\ref{tab:topics} in the 13 courses of Table~\ref{tab:existing-courses}. On the x-axis there is the label corresponding to the topic of Table~\ref{tab:topics}. On the y-axis there is the number of courses (those reported in Table~\ref{tab:existing-courses}) that consider the topic. Topic 3 (Classification), 4 (Controls), and 10 (Information theory) are never considered by any course. The topics treated with highest popularity are 2 (Applications), 18 (Routing), 12 (MAC) and 16 (Programming).   }
\label{fig:frequency-course-topics}
\end{figure}

In Figure~\ref{fig:frequency-course-topics}, the occurrence of the topics of Table~\ref{tab:topics} in the courses of Table \ref{tab:existing-courses} is quantified. 
The topics taught with highest popularity are 2 (Applications), 18 (Routing), 12 (MAC) and 16 (Programming). Obviously, 2 is the most popular topic, since applications give a strong motivation to study WSNs.  Topic 3 (Classification), and 4 (Controls) are never considered by any existing course. 

\begin{figure}[t]
\centering
\includegraphics[width=9cm]{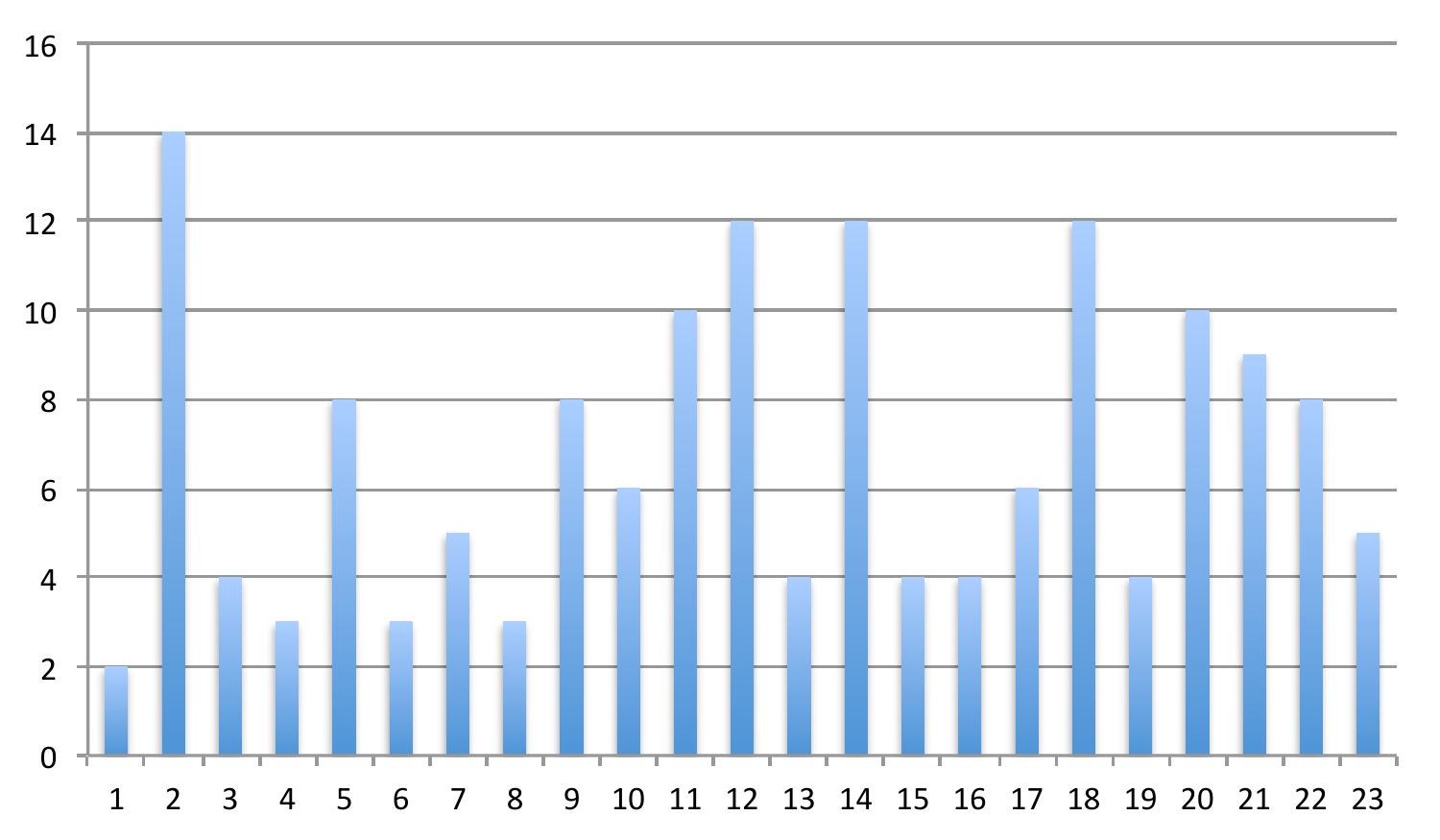}
\caption{Number of occurrences of a topic of Table~\ref{tab:topics} in the 14 books~\cite{ZhaoGuibas04}--\cite{WagnerWattenhofer10}. On the x-axis there is the label corresponding to the topic of Table~\ref{tab:topics}. On the y-axis there is the number of books (reported in \cite{ZhaoGuibas04}--\cite{WagnerWattenhofer10}) that consider the topic. The topics having the highest popularity are 2 (Applications), 12 (MAC), 14 (Network management), and 18 (Routing). The topics with the lowest popularity are 1 (Antennas), 4 (Controls), 6 (Detection), and 19 (Sensor principles).}
\label{fig:frequency-book-topics}
\end{figure}

In Figure~\ref{fig:frequency-book-topics}, the occurrence of the topics of Table~\ref{tab:topics} in the existing books is reported. The topics described with highest popularity are 2 (Applications), 12 (MAC), 14 (Network management), and 18 (Routing). As for the courses, 2 is the most popular topic. The rest of the popular topics is due to that there has been considerable research in the Networking communities of both Electrical Engineering and Computer Sciences to define new MAC and routing protocols and the network management especially to save energy consumption. The topics with the lowest popularity are 1 (Antennas), 4 (Controls), 6 (Detection), and 19 (Sensor principles). The low popularity of Antennas and Sensor principles is of no surprise, since these topics are somewhat considered traditional and a part of sensor systems, namely system where there are sensors, but which are not necessarily connected to form a network. A similar observation holds for the topic Detection as wel, which is a traditional topic having even stand-alone monographs for sensor systems, e.g.,~\cite{Varshney97,Gustafsson12}. By contrast, the low popularity of controls topics is remarkable, because control of WSNs is an essential design tool and control over WSNs is a pervasive application, especially in industrial and building automation, as remarked in~\cite{Rollins11}.  

Now that we have identified the essential knowledge needed for WSNs, we are now in the position of proposing a course design in the next section.

\section{Course design} \label{course design}

In this section, we propose a course content for an 8 weeks course of 60 hours in classroom, approximately 8 hours per week in classroom, plus 30 hours homework, as normal for Master courses at KTH Royal Institute of Technology, Sweden, for Electrical Engineering and Computer Sciences students. The course proposal includes the lecture content, the homework, project, and exam. In the following, we illustrate the method we have followed, and then the course details. 

\subsection{Design methodology}

The literature survey analyzed in Section~\ref{literature} and Section~\ref{relevant knowledge}, including books and existing course worldwide, were considered to select the most relevant topics. Afterwards, the course design is based on meetings and questionnaires obtained with interviews of industrial representatives, academic representatives and students. Specifically, the following world-leading companies were considered, where we have specified the area of interest:
Ericsson Research, South Korea (Machine-to-Machine and Device-to-Device communications); Ericsson Research, Sweden (4/5G Communications, Machine-to-Machine and Device-to-Device communications); Ericsson Business Units, Sweden (4/5G Communications, Machine-to-Machine and Device-to-Device communications);
ABB Corporate research, Sweden and Norway (Industrial factory and process automation); Electrolux, Sweden (Building automation); Uniter Technology Research Center, Connecticut and California, USA (Industrial factory and process automation, Building automation); Fortum AB, Sweden (Smart grids); Telecom Italia, Italy (4/5G Communications, Machine-to-Machine and Device-to-Device communications); Terna, Italy (Smart grids); Acciona Agua, Spain (Industrial process automation);
Ottobocke, Austria (Health care); Thales, France (Military); Fiat, Italy (Vehicular communications); Cisco, Switzerland (Industrial automation);

Within academic institution, we have asked the following: KTH Royal Institute of Technology, representatives of the Master Program in System Controls and Robotics and Wireless Communications; University of L'Aquila, Italy; University of Padova, Italy; McGill University, Canada; University of California at Berkeley, USA; 
University of California at Irvine, USA; University of Lund, Sweden; Polytechnic Institute of Porto, Portugal; KAIST, Korea; University of Valencia, Spain; FORTH Center, Greece; University of Seville, Spain; Massachusetts Institute of Technology, USA; Stanford University, USA.
 
Four different questionnaires were carried out with the following purposes:
\begin{enumerate}
\item Asking to the industrial representatives listed above which topics of Table~\ref{tab:topics} the course should offer.
\item Asking to academic colleagues of the institutions listed above which topics of Table~\ref{tab:topics} the course should offer.
\item	 Asking to the relevant KTH Master Program representatives how to harmonize the course content with other existing courses covering the topics of topics of Table~\ref{tab:topics} in the Master program. 
\end{enumerate}

The course design therefore reflects thoroughly the data collected, including bibliography, existing courses worldwide, interviews of teachers and students, and industrial representatives. 

\subsection{General course description} \label{course-description}

The course is an 8 weeks course of 60 hours in classroom plus 30 hours homework to be taught at the fourth year or five year out of five years Engineering program (3 years Bachelor level and 2 years Master level). Prerequisites include courses such as ``Signal and Systems'' and ``Probability''. The learning outcome consists in 1) knowing the essential Networking, Signal Processing, and Controls to cope with WSNs, 2) knowing how to design practical WSNs, and 3) be able to develop a research project on WSNs. 

The course is based on a both theoretical and hands-on approach, where the theoretical aspects are first taught in lectures, and then they are better mastered in exercise sessions, where the concepts are illustrated by exercises with questions, and by experimental implementations. These experiments are not just to let the students making an experience out of the theory, but actually to show how the theory is useful to design practical real world engineering systems. 

In the following, we give the course content description along with the intended learning outcomes, the homework and exam description.

\subsection{Course content} \label{course-content}

\begin{table}[t]
\tiny
\begin{center}
\caption{Proposal for a course. The numbers on the right refers to Table~\ref{tab:topics}}
  \begin{tabular}{||  l || l  || l  ||  }
   \hline Lecture & Title & Topics \\
   \hline
    \hline 1 & Introduction &  1, 2, 7, 18,  21\\
     \hline 1 & Exercises  &  \\
    \hline 2 & WSNs hardware and programming & 14, 15 \\
    \hline 2 &  Exercises & \\
     \hline
     \hline PART 1  & Networking &   \\
       \hline
    \hline 3 & Wireless channel & 16 \\
    \hline 3 & Exercises & \\
    \hline 4 & Physical layer  & 9, 12, 19\\
     \hline 4 & Exercises  & \\
    \hline 5 & MAC Layer &  11, 22\\
    \hline 5 & Exercises &  \\
    \hline 6 & Routing  &  17, 22\\
     \hline 6 & Exercises  &  \\
      \hline
     \hline   & HOMEWORK 1 &   \\
       \hline
    \hline
     \hline PART 2  & Signal processing &   \\
       \hline
    \hline 7 & Distributed detection &  3, 6\\
      \hline 7 & Exercises &  \\
    \hline 8 & Distributed static estimation & 3, 6 \\
      \hline 8 & Exercises &  \\
       \hline 9 & Distributed dynamic estimation & 3, 6 \\
      \hline 9 & Exercises &  \\
    \hline 10 & Localization and positioning  & 10 \\
      \hline 10 & Exercises &  \\
    \hline 11 & Time synchronization  &  20 \\
      \hline 11 & Exercises  &  \\
      \hline
     \hline   & HOMEWORK 2 &   \\
       \hline
      \hline
     \hline PART 3 & Controls &   \\
       \hline
    \hline 12 & WSNetworked control 1 & 4 \\
      \hline 12 & Exercises &  \\
    \hline 13 & WSNetworked control 2  &  4 \\
      \hline 13 & Exercises &  \\
    \hline 14 & Network control  &  4, 5 \\
      \hline 14 & Exercises &  \\
    \hline 15 & Course summary & 1--22 \\
      \hline 15 & Exercises, example of exam &  \\
      \hline
     \hline   & HOMEWORK 3 &   \\
       \hline
    \hline
  \end{tabular}
  \label{tab:course-proposal}
\end{center}
\end{table}

In order to master the interdisciplinarity of WSNs, the course is divided into three parts after two introductory lectures, one on WSNs applications and one on software programming. The three parts are organized as follows: Networking, Signal Processing, and Controls, because every part builds on the previous one. For every part, the most relevant topics are selected based on the industry needs, popularity in existing courses, books, and personal experience, as we have mentioned in the previous subsections. Moreover, the selection of the topics is done so as to give an holistic view on WSNs. 

More specifically, the course is structured as reported in Table~\ref{tab:course-proposal}.  Per every lecture of the table, the intended learning outcomes answer the following questions: 
\begin{itemize}
\item Lecture 1: What are the components of a WSN? What are typical applications of a WSN? What is a networking protocol? How to design applications and protocols?
\item Lecture 2: What are the operating systems that are available? How to program WSNs? 
\item Lecture 3: How bits of information are transmitted (digitally modulated) over a wireless channel? What is the AWGN channel? How the wireless channel attenuates the transmit radio power? 
\item Lecture 4: What is the probability to receive correctly messages over AWGN
channels as function of the transmit power? What is the probability to receive correctly messages over fading
channels?
\item Lecture 5: What is the Medium Access Control (MAC)? How the modulation and channel influences the MAC performance? What are the options to design MACs?
What is the MAC of the important communication standard IEEE 802.15.4?
\item Lecture 6: What are the basic routing options? How routing and MAC are connected? How to compute the shortest path from a source to a destination? Which routing is used in standard protocols?
\item Lecture 7: How to detect events? What is the probability of miss detection and false alarm? How Networking impact these procedures? 
\item Lecture 8: How to estimate a random variable? How to estimate from static measurements over a network? How Networking has an impact on estimation? 
\item Lecture 9: How to perform distributed dynamic estimation from noisy measurements? How Networking has an impact on distributed estimation? 
\item Lecture 10: Which measurements are used for estimating the position of a node? How to estimate the position of a node by a network of sensors?
\item Lecture 11: Which measurements are used for synchronizing the nodes? What is the hardware and the software clock? How to synchronize the nodes in a centralized manner, and in a distributed?
\item Lecture 12: How to model mathematically a wireless sensor networked control system? What is a typical proportional-inegrator-derivative controller? How a closed loop control system with periodic sampling is affected by delays and packet losses in the communication?
\item Lecture 13: How to jointly design the WSN protocols (modulations, MAC, routing) estimators and controllers?
\item Lecture 14: How to optimize all the layers of the WSNs protocols and applications on top of the network? How to control the network topology?
\item Lecture 15: Course summary. How the final exam is structured? 
\end{itemize}

The learning outcomes per every theoretical lecture is reinforced during the exercises lecture, where the student have hands-on an experimental test bed so to see how to design in practice a WSN based on theory. 

We now turn our attention to the design of homework. 

\subsection{Homework design} \label{subsection:homework}

The course gives three homeworks for the student, one per each part of the course content: Networking, Signal Processing, and Controls. Every homework contains partly experimental and partly theoretical questions. The theoretical questions are related to problems that can be solved by using the methods taught during the lectures, whereas the experimental part is used to illustrate and better understand the theoretical concepts. 
The homework can be done by a team. It is recommended that no more than three students should participate in a single team, which in turn can increase the individual involvement and thus maximize the learning potentials. In addition, five sensors per homework per students are recommended, where few sensors are connected to a physical phenomena and other sensors act as relay in a multi hop routing tree toward a sink node. The homework's are described in the following: 

\begin{enumerate}
\item Homework 1: it covers the lectures 2--6. The students are asked to solve theoretical problems that are useful for real-world implementation concerning networking aspects, such as computing the probability of error in the reception of messages, how the modulation formats and coding influence such a probability. A question to compare between the medium access control protocols based on CSMA and TDMA is proposed. The effect of radio power, modulations, packet sizes, and MAC options are studied in the routing protocol. An experimental implementation of the networking aspects over a real-world WSN test bed has to be performed. 
\item Homework 2: it covers the lectures 7--11. The students are asked to solve theoretical problems concerning how to detect events by WSNs, how to estimate the characteristics of those events, how to localize the position of nodes and events, and finally, how to synchronize the clock of the nodes. The experimental implementation covers these topics one by one, where the students have to characterize experimentally the performance of the various detection and estimation methods. 
\item Homework 3: it covers the lectures 12--14 in the detail, but also the rest of the course. The Controls part of the course puts together all the previous topics of the course and hence is arguably the most complex part. Therefore, this homework has a reduced theoretical part, and is mostly focused on experiments. In particular, the students are requested to implement a PID controller, which is typically used in building and factory automation, where the process is running on a computer connected to a sensor, and the controller on another computer connected to another sensors. The communication between the two computers is done via a WSN to control the process. The effect of the packet losses, MAC, and routing on the control performance is studied by experimental activities. 
\end{enumerate}

The homeworks do not account for the final grading of the course, but they have a separate pass/fail grade. When the homeworks are completed, the corresponding students are entitled to sit for the exam. The main purpose of the homework is to maximize the learning, to increase as much as possible knowledge retaining, and to stimulate the creativity especially with the experimental activities. 

Finally, let us turn our attention to the exam design. 

\subsection{Exam design} \label{exam design}

The purpose of the exam is to give the students a further learning opportunity by seeing all the course content together, and also to give a psychological motivation to study. 

The exam contains only theoretical problems, which are composed on the lines of those of the exercise sessions and of the homework. However, the problems of the exam will cover the course content in a unified way, so that there is a further chance for the students to learn how all the pieces of the course are interconnected.  

Five problems compose the exam. Two on the first part of the course, Networking, two on the second part, Signal Processing, and one on the last part, Controls. Every problem accounts for 10 points, and thus the final grade sums up to 50 points. Then, according to the KTH grading scale, A corresponds to more than 43 points, B corresponds to points between 38 and 42, C corresponds to points between 33 and 37, D corresponds to points between 28 and 32, and E corresponds to points between 23 and 27. Below 23, the exam is failed.  

The students do not have to gain all the points from the exam. They are given the possibility to develop a Hands-on Project where they can gain up to 15 points, as explained below, and thus avoid doing all the exercises at the exam. However, the sum of the points of the project and of the exam problems saturates to $50$. 

\subsection{Hands-on Project}\label{handsonprojects}

The students can choose to develop a project mostly consisting on experimental activities. The project is supposed to cover a part of the course corresponding to one or two lectures where the experiments have to show in detail the theoretical implications of physical causes of the chosen topic. The students are given the freedom of choosing the lecture associated to the project. For example, in the case of Lecture 3, experiments have to be performed to characterize the distribution of the wireless channel in many sensor networks environments and for many situations of fading. 

The project has to be described in a report having the IEEE format in double column and in an oral presentation. It accounts to 15 points, which can be summed up to the points achieved by the final exam. 

\section{Evaluation of the first and second edition of the course} 

The course is evaluated based on two editions, in 2012 and 2013. 

The first edition of the course was given in the Winter term 2012. This 2012 course had a similar content to Table~\ref{tab:course-proposal}, with in addition the topics of Table~\ref{tab:topics} 19, Sensor Principles, and 22, Security and Privacy. The 2012 course did not have experimental implementations in all homeworks, conversely to what proposed in Subsection~\ref{subsection:homework}, but only for the first homework. Moreover, the 2012 course did not have any hands on project. The course was taken by 15 students. The final course evaluation asked to evaluate the course among ``bad'', ``sufficient'', ``good'', ``very good'', ``excellent'', and it showed that $50\%$ of the students considered the course ``good'' or ``very good''. Most of the students were not interested to have Sensor Principles and Security topics (see Subsection~\ref{subsec:topic-classific} for the description of these topics) because they considered that these were extrinsic topics compared to the rest of the course contents. Moreover, there was an overwhelming consensus that the experimental part was very appealing and should have been extended. 

The second edition of the course was given in the Winter 2013. The course is the one described in the previous Subsections~\ref{course-description},~\ref{course-content},~\ref{subsection:homework},~\ref{exam design}, and~\ref{handsonprojects}. Compared to the 2012 edition, the two lectures on Sensor Principles and on Security and Privacy were removed (as suggested by the students of the first edition). More experimental activities and hands-on projects were introduced. The course was taken by $25$ students. At this time, $75\%$ of the students rated the course either ``very good'' or ``excellent''. Moreover,  $82\%$ of the students thought that the hands-on project was very useful and enjoyable. A small fraction suggested replacing entirely the exam by an experimental project. While this is certainly appealing and enjoyable, we believe it would not allow the students to master the many concepts from different disciplines being considered in the course. Experimental projects need necessarily to be focused on some aspect, which would neglect other important ones. It is worth noting that theory and practice are equally important in this course. The practical part is already substantially considered in the exercise lectures and in the homework. The student evaluation suggested that not all the students of the 3-rd year may have the necessary background in Networking, Signal Processing, and Controls. The evaluation thus suggested that the course could be given both at the 4-th year and the 5-th year. We believe, this would allow students  to be equipped with useful mathematical tools and to become more matured with other related courses, which can potentially enhance the learning potentials for WSNs. As a consequence, the 2014 edition will be open also to the 5-th year students. Finally, the students complained about the absence of a book that could present the material in a unified manner. They suffered from that the material was taken from different sources whose heterogeneous style, language, and mathematical symbols was a barrier to a more efficient learning. In my opinion, this observation supports the need of a development of a more complete book. 

\section{Conclusion and Future Developments} \label{conclusions}

This paper provided a first approach toward a characterization of the the basic knowledge for WSNs and effective teaching and learning methods at the Master level for a curriculum in Electrical Engineering and Computer Sciences. The analysis was based on the review of existing courses worldwide, books, research papers, and questionnaires to industry and academia. It was shown that WSNs require basic knowledge from Networking, Signal Processing, Controls and Embedded programming, and thus the necessary topics from these areas of study were hereby individuated. A proposal for a course that includes such a knowledge as well as relevant lab experiences, exercises, and homework was finally given, where the focus is on both a theoretical and an experimental implementation. The course was evaluated over two years. It was observed that the proposed holistic approach can be highly effective for teaching and learning of WSNs in both Electrical Engineering and Computer Sciences. 

We believe that the WSNs learning can be greatly improved by the suggested course. Given the interdisciplinary of the course, students can then deepen their interest in several directions in the future with great benefit for their formative growth. 

\section{Acknowledgmenet}

We are grateful to Khalid El Gaidi of KTH Learning Lab for numerous and inspiring discussions on the paper content, and to Euhanna Gadimi and Yuzhe Xu for many useful discussions especially on the exercises and homework. We are grateful to George Athanasiou, Piergiuseppe Di Marco, Marco Levorato, and Chathuranga Weeraddana for providing useful feedback. 

This work was supported by the EU projects Hydrobionets and Hycon2. 

\bibliographystyle{IEEEtran}
\bibliography{ref}

\end{document}